Comparison of Transcriptional Activation by Corticosteroids of Human MR (Ile-180) and Human MR Haplotype (Ile180Val)

Yoshinao Katsu[1,2,*], Jiawen Zhang[2], Ya Ao[2], Michael E. Baker[3,4,*]

[1] Faculty of Science

Hokkaido University

Sapporo, Japan

[2] Graduate School of Life Science

Hokkaido University

Sapporo, Japan

[3] Division of Nephrology-Hypertension

Department of Medicine, 0693

University of California, San Diego

9500 Gilman Drive

La Jolla, CA 92093-0693

Center for Academic Research and Training in Anthropogeny (CARTA) [4]

University of California, San Diego

La Jolla, CA 92093

*Correspondence to

Y. Katsu; E-mail: ykatsu@sci.hokudai.ac.jp (YK)

M. E. Baker; E-mail: mbaker@health.ucsd.edu (MEB)

**Abstract** While the classical function of human mineralocorticoid receptor (MR) is to regulate sodium and potassium homeostasis through aldosterone activation of the kidney MR, the MR also is highly expressed in the brain, where the MR is activated by cortisol in response to stress. Here, we report the half-maximal response (EC50) and fold-activation by cortisol, aldosterone and other corticosteroids of human MR rs5522, a haplotype containing valine at codon 180 instead of isoleucine found in the wild-type MR (Ile-180). MR rs5522 (Val-180) has been studied for its actions in the human brain involving coping with stress and depression. We compared the EC50 and fold-activation by corticosteroids of MR rs5522 and wild-type MR transfected into HEK293 cells with either the TAT3 promoter or the MMTV promoter. Parallel studies investigated the binding of MR antagonists, spironolactone and progesterone, to MR rs5522. In HEK293 cells with the MMTV promotor, MR rs5522 had a slightly higher EC50 compared to wild-type MR and a similar level of fold-activation for all corticosteroids. In contrast, in HEK293 cells with the TAT3 promoter, MR 5522 had a higher EC50 (lower affinity) and higher fold-activation for cortisol compared to wild-type MR (Ile-180), while compared to wild-type MR, the EC50s of MR rs5522 for aldosterone and corticosterone were slightly lower and fold-activation was higher. Spironolactone and progesterone had similar antagonist activity for MR rs5522 and MR (Ile-180) in the presence of MMTV and TAT3 promoters in HEK293 cells.

## Introduction

The mineralocorticoid receptor (MR) is a ligand-activated transcription factor [1] that belongs to the nuclear receptor family along with other vertebrate steroid receptors: the glucocorticoid receptor (GR), progesterone receptor (PR), androgen receptor (AR) and estrogen receptor (ER) [1–6]. In humans and other terrestrial vertebrates the MR maintains electrolyte balance by regulating sodium and potassium transport in epithelial cells in the kidney and colon [7–12]. However, the MR also has important physiological functions in many other tissues, including brain, heart, skin and lungs [11,13–24].

Analysis of the human MR sequence by Arriza et al [1] revealed that, like other steroid receptors, the human MR is composed of four modular functional domains: a large amino-terminal domain (NTD) of about 600 amino acids, followed in the center by a DNA-binding domain (DBD) of about 65 amino acids, followed by a small hinge domain of about 60 amino acids that is connected to the ligand-binding domain of about 250 amino acids at the C-terminus, where aldosterone, cortisol and other corticosteroids bind to activate transcription [1,2,25–27].

Arriza et al. [1] also reported that the MR sequence was closely related to the glucocorticoid receptor (GR) sequence, consistent with evidence that some corticosteroids, such as cortisol and corticosterone were ligands for both the MR and GR [13,15,20,26,28] and that aldosterone, cortisol, corticosterone and 11-deoxycorticosterone have similar binding affinity for human MR [1,20,28–30]. Activation of the MR by cortisol and corticosterone, two steroids that are ligands for the GR [30–32], is consistent with the evolution of the GR and MR from a common ancestral corticoid receptor (CR) in a cyclostome (jawless fish) that evolved about 550 million years ago at the base of the vertebrate line [26,33–39].

Humans contain three almost identical MR transcripts that differ at codons 180 and 241 in the NTD [40] (Fig 1). Human MR (Ile-180), cloned by Arriza et al [1], has been extensively studied for transcriptional activation by aldosterone and other corticosteroids [1,28,29,31,32,40]. Transcriptional activation by aldosterone and other corticosteroids of human MR (Ile-180, Val-241) and MR (Val-180, Val-241) has recently been reported [40]. Here, we focus on the response to corticosteroids of another human MR, haplotype rs5522 (Val-180), which contains a valine at codon 180 [41–44] (Figure 1). rs5522 (Val-180) has important physiological effects on stress and depression in humans [43–47] making transcriptional activation by steroids of rs5522 of much interest.

Aldosterone [42,43] and cortisol [43,48] are transcriptional activators of MR rs5522. However, EC50s and fold-activation by aldosterone, corticosterone and 11-deoxycorticosterone for the rs5522 haplotype have not been reported. Activation of rs5522 by a corticosteroid other than cortisol could provide clues to the unique physiological actions of rs5522 in the brain MR.

Nor has the binding of spironolactone and progesterone, two MR antagonists [49,50] to rs5522been investigated.. To provide this information, we investigated the effect of corticosteroids and of mineralocorticoid antagonists on transcriptional activation of rs5522 in presence of two promotors: TAT3 [51,52] and MMTV [25,53]. We find that in HEK293 cells with the MMTV promotor, each corticosteroid had a similar EC50 and fold-activation for rs5522 and wild-type MR (MR Ile-180). In contrast, in HEK293 cells with the TAT3 promoter, compared to wild-type MR, rs5522 had a slightly lower EC50 for aldosterone and corticosterone and a higher EC50 for cortisol and a similar EC50 for 11-deoxycorticosterone. However, fold-activation for cortisol activation of rs5522 was 1.4-fold higher than cortisol activation of MR (Ile-180), which is interesting because cortisol is the biological ligand for the brain MR [17,18,20,22,23]. Compared to wild-type MR (Ile-180), rs5522 had a higher fold-activation of 1.26, 1.29 and 1.32, respectively, for aldosterone, corticosterone and 11-deoxycorticosterone. We find that spironolactone and progesterone had similar antagonist activity for rs5522 and MR (Ile-180) in the presence of the MMTV and TAT3 promoters in HEK293 cells. The absence of a large difference in EC50 and/or fold-activation between rs5522 and (MR Ile-180) in their response to corticosteroids suggests that other mechanisms are important in rs5522 regulation of the response to stress and depression. These may include post-translational modification of the NTD [24,54–58], formation of MR-GR heterodimers [59–64] and binding to specific promoters [61,65].

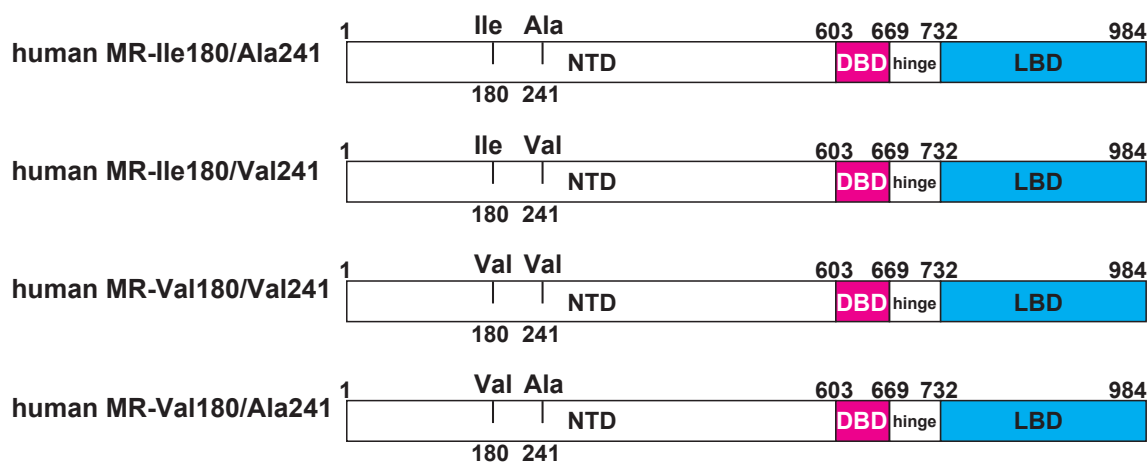

**Fig 1. Human MR Genes.**

There are three human MR genes with 984 amino acids in GenBank. These MRs contain either isoleucine-180, alanine-241 (Ile180/Ala241: accession AAA59571), isoleucine-180, valine-241 (Ile180/Val241: accession XP_054206038), or valine-180, valine-241 (Val180/Val241: accession NP_000892), in the NTD. The fourth MR isoform is rs5522, which contains valine-180 instead of isoleucine-180 (Val180/Ala241) [41–43].

Results

**Corticosteroid-dependent activation of wild-type human MR (Ile-180) and rs5522 (MR Val-180)**

We have used human MR (accession AAA59571) [1], which contains isoleucine at codon 180 for our studies of the human MR [29,40]. For this project, as described in the Methods Section, we constructed rs5522 from human MR (Ile-180, Ala-241). In Fig 2, we show the concentration dependence of transcriptional activation by corticosteroids of rs5522 and MR (Ile-180) transfected into HEK293 cells with either an MMTV or a TAT3 luciferase promoter. Luciferase levels were used to calculate an EC50 and fold-activation for each steroid (Table 1).

Fig 2 reveals that in HEK293 cells, transfected with the MMTV promoter, the EC50 and fold-activation values were similar for aldosterone, cortisol, corticosterone and 11-deoxycorticosterone. It is in the experiments with TAT3 promoter in HEK293 cells that we find differences between rs5522 and MR (Ile-180). As shown in Fig 2 and Table 1, in HEK293 cells transfected with TAT3, and with either aldosterone or corticosterone had a slight decrease in EC50 (stronger binding) for rs5522 compared to MR (Ile-180), while cortisol had a slight increase in EC50 for rs5522 compared to MR (Ile-180). The EC50 of 11-deoxycorticosterone was about the same value for rs5522 and MR (Ile-180).

Also as shown in Fig 2 and Table 1, in HEK293 cells transfected with TAT3, aldosterone increased fold-activation by about 25% for rs5522 compared to MR (Ile-180). Cortisol, which activates the MR in organs that lack 11β-hydroxysteroid dehydrogenase-type 2 [7,66–68], increases fold activation by about 40% for rs5522 compared to MR (Ile-180). Corticosterone and 11-deoxycorticosterone increased fold-activation by about 30% for rs5522 compared to MR (Ile-180).

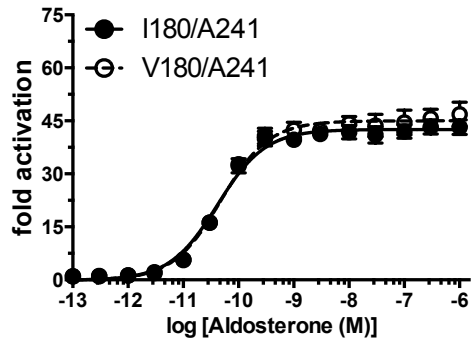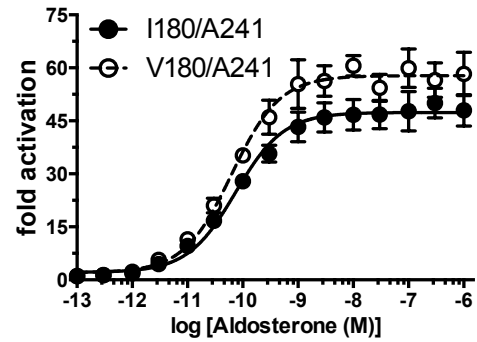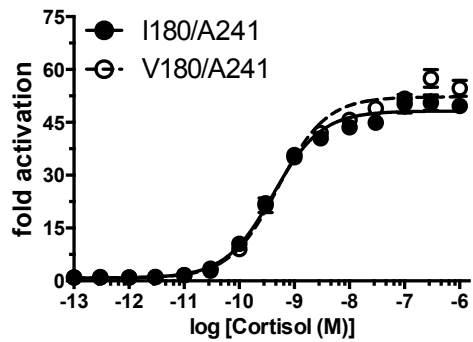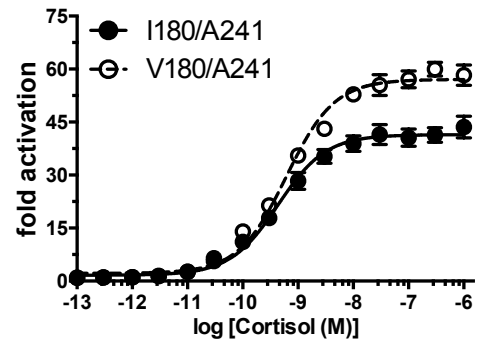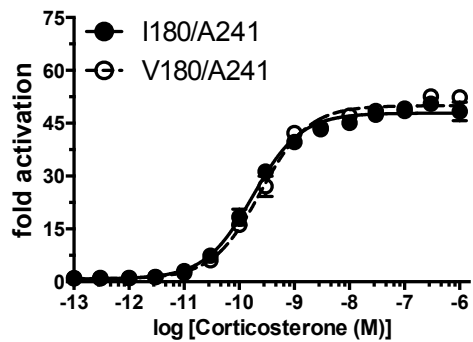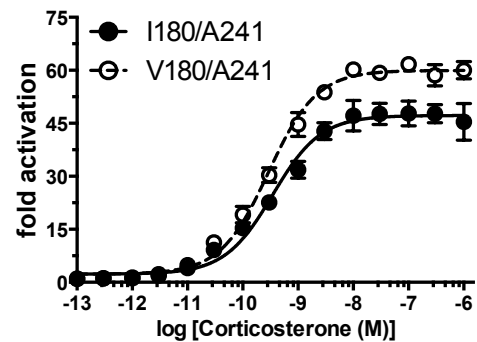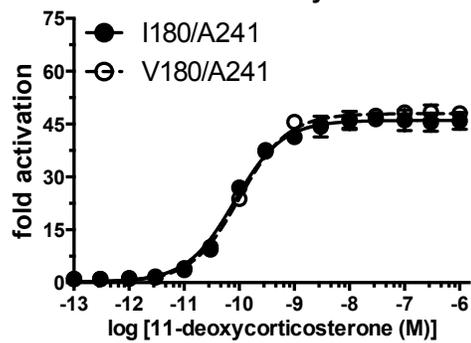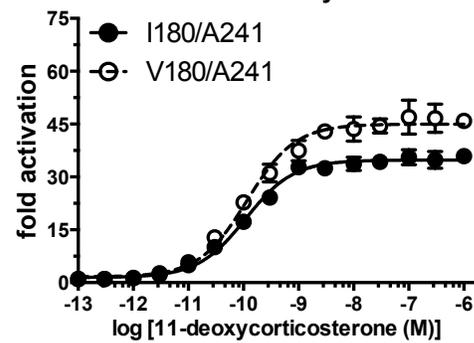

**Fig 2. Concentration-dependent transcriptional activation by corticosteroids of wild-type human MR (Ile-180) and rs5522 (MR Val-180) in the presence of either the MMTV promoter or the TAT3 promoter.**

Plasmids for wild-type MR (Ile-180) and (MR Val-180) were expressed in HEK293 cells with either an MMTV-luciferase promoter (2A-2D) [25,53] or a TAT3-luciferase promoter (2E-2H) [2,51,52]. Cells were treated with increasing concentrations of either aldosterone, cortisol, 11-deoxycorticosterone, corticosterone, or vehicle alone (DMSO). Results are expressed as means ± SEM, n=3. Y-axis indicates fold-activation compared to the activity of vector with vehicle (DMSO) alone as 1. A. MR (Ile-180) and (MR Val-180) with aldosterone. B. MR (Ile-180) and (MR Val-180) with cortisol. C. MR (Ile-180) and (MR Val-180) with corticosterone. D. MR (Ile-180) and (MR Val-180) with 11-deoxycorticosterone. E. MR (Ile-180) and rs5522 (MR Val-180) with aldosterone. F. MR (Ile-180) and rs5522 (MR Val-180) with cortisol. G. MR (Ile-180) and (MR Val-180) with corticosterone. H. MR (Ile-180) and (MR Val-180) with 11-deoxycorticosterone.

**Table 1. Steroid Activation of Human MRs in HEK293 Cells with an MMTV or a TAT3 Promoter.**

|  |  | MMTV-luc | | TAT3-luc | |
| --- | --- | --- | --- | --- | --- |
|  |  | I180/A241 | V180/A241 | I180/A241 | V180/A241 |
| Aldosterone | EC50 (nM) | 0.042 | 0.048 | 0.074 | 0.063 |
|  | Fold-Activation (± SEM) * | 41.9 (± 1.9) | 44.5 (± 3.5) | 47.7 (± 5.6) | 59.9 (± 5.4) |
|  | Ratio** | 1 | 1.06 | 1 | 1.26 |
| Cortisol | EC50 (nM) | 0.41 | 0.51 | 0.43 | 0.62 |
|  | Fold-Activation (± SEM) * | 50.3 (± 2.6) | 51.7 (± 0.9) | 40.6 (± 2.4) | 57.1 (± 2.4) |
|  | Ratio** | 1 | 1.03 | 1 | 1.41 |
| Corticosterone | EC50 (nM) | 0.18 | 0.24 | 0.35 | 0.29 |
|  | Fold-Activation (± SEM) * | 48.5 (± 1.7) | 49.2 (± 0.9) | 47.8 (± 3.5) | 61.7 (± 0.06) |
|  | Ratio** | 1 | 1.01 | 1 | 1.29 |
| 11-deoxy-corticosterone | EC50 (nM) | 0.083 | 0.1 | 0.11 | 0.12 |
|  | Fold-Activation (± SEM) * | 46.1 (± 3.0) | 48.3 (± 1.2) | 35.6 (± 2.1) | 47.0 (± 4.8) |
|  | Ratio** | 1 | 1.05 | 1 | 1.32 |

*Fold activation values were calculated based on the 100 nM values.

**Ratio values were calculated by dividing the fold-activation value of V180/A241 form by the value of I180/A241 form.

**Spironolactone and progesterone are antagonists for rs5522 and MR (Ile-180)**

To determine if spironolactone and progesterone are antagonists for rs5522, we added either spironolactone or progesterone to HEK293 cells transfected with either human MR (Ile-180) or rs5522 and either an MMTV promoter or a TAT3 promoter and incubated with 0.1 nM aldosterone. Fig 3 shows that 10 nM spironolactone or 10 nM progesterone inhibit activation of rs5522 and MR (Ile-180) by 0.1 nM aldosterone. We conclude that spironolactone and progesterone are antagonists of rs5522 [49,50].

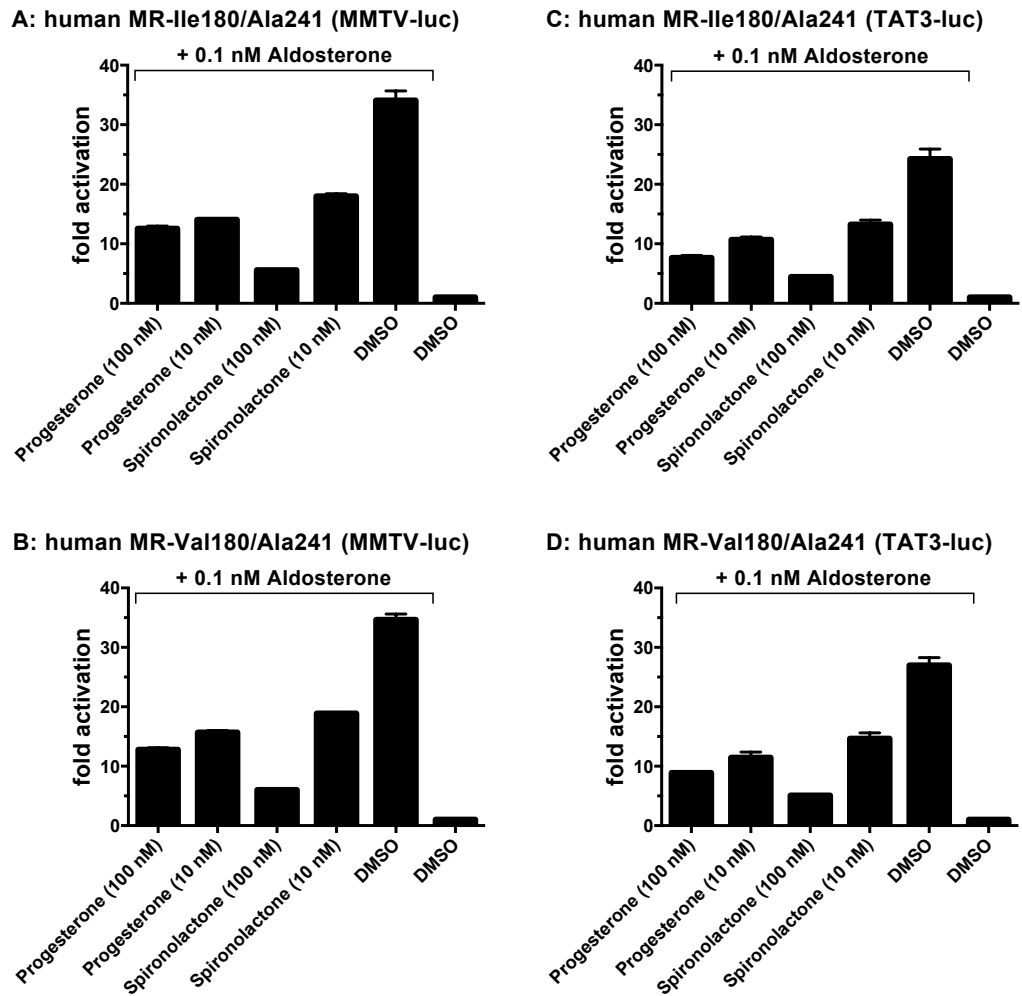

**Fig 3. Inhibition of Aldosterone Activation rs5522 and human MR (Ile-180, Ala-241) by spironolactone and progesterone.**

**Figure Legend:** HEK293 cells were transfected with either human MR (Ile-180, Ala-241) or rs5522 (Val-180, Ala-241) and either the MMTV promoter or the TAT3 promoter. Then these cells were incubated with 0.1 nM aldosterone alone or with either 10 nM or 100 nM spironolactone or progesterone for 24 hrs. HEK293 cells were harvested and processed for luciferase.

**FLAG-tagged expression of human MR (Ile-180) and rs5522**

To investigate the protein expression levels of FLAG-tagged MR (Ile-180) and rs5522 (Val-180), HEK293 cells transfected with each construct were collected and treated with SDS sample buffer. Subsequently, FLAG-tagged proteins were subjected to SDS-PAGE on a 10% polyacrylamide gel and transferred to an Immobilon membrane and detected using an antibody against the FLAG tag. We detected a single band of human MR (Ile-180) and rs5522 (Fig 4), which were used in the experiments depicted in Fig 2. This analysis of FLAG-tagged human MR (Ile-180) and rs5522 (Val-180) confirms that there were similar protein expression levels of these two MRs in our experiments.

**Fig 4 FLAG-tagged expression of human MR (Ile-180) rs5522 (Val-180).**
HEK293 cell lysates transfected with FLAG-tagged MR (Ile-180, Ala-241) and rs5522 (Val-180, Ala-241) were treated with sample buffer and applied to a 10% SDS-polyacrylamide gel, and

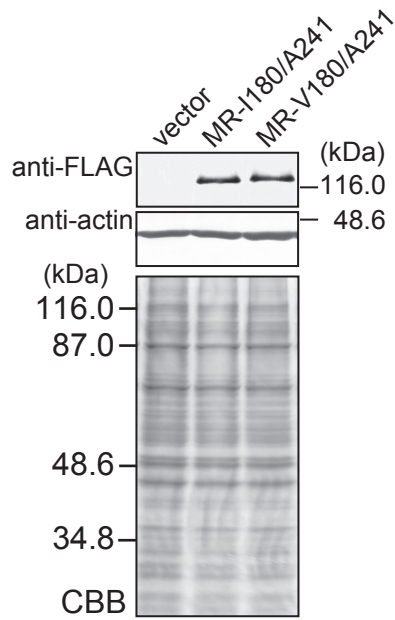

then transferred onto a membrane. The expressed FLAG-tagged proteins were detected with anti-FLAG antibody. The expression level of endogenous actin was measured using an actin antibody. Upper panel: Western blot. Lower panel: Coomassie blue stain (CBB).

**Discussion**

The cloning of the human MR by Arriza et al [1] provided tools for investigating the function of the MR in diverse tissues other than the distal tubule of the kidney, including the MR in the hippocampus, which is receiving increasing attention as an MR-responsive organ [19,20,22,23]. Indeed, the MR in the hippocampus is downregulated by extended stress and depression [20,45,69,70], and MR synthesis is induced by anti-depressants. Increased activity of the MR inhibits activity of the hypothalamic-pituitary-adrenal axis and promotes slow wave cognition, reducing anxiety.

The MR rs5522 (Val-180), which is the focus of this report, is intriguing and important because humans with MR (Val-180) have increased response to stress [22,23,43,46] despite the presence of only a single amino acid difference between the rs5522, which contains valine at codon 180, instead of isoleucine at codon 180 as found in wild-type MR [22,45,46,71]. Such a profound physiological effect of a point mutation in the NTD, which is distant from the LBD and DBD is unexpected. Moreover, valine-180 is not located in one of the transcriptional activation domains in the NTD, which occur at residues 1-169 and 451-603 in the NTD of human MR [72,73].

As a first step towards understanding the biological actions of the Ile180Val mutation in human MR, we investigated transcriptional activation of MR (Val-180) by a panel of corticosteroids including aldosterone and cortisol, for comparison with activation of wild-type human MR [Figure 2, Table 1]. We studied transcriptional activation in the presence of either the MMTV promoter or the TAT3 promoter. Interestingly, in cells with TAT3, but not in cells with MMTV, there were differences in transcriptional activation of MR (Val-180) and wild-type MR (Ile-180) by corticosteroids.

Among the corticosteroids, cortisol is of interest because in the human brain cortisol is the physiological steroid for the MR because the circulating cortisol concentration is about 200-fold higher compared to aldosterone, and 11β-hydroxysteroid dehydrogenase-type 2, which can inactivate cortisol [66–68], is absent in the hippocampus [66]. Cortisol had a higher EC50 for MR (Val-180) than for wild-type MR in HEK293 cells transfected with TAT3, in agreement with [43]. Interestingly, fold-activation of MR (Val-180)1 by cortisol was about 40% higher than for cortisol activation of wild-type MR. Aldosterone had a slightly lower EC50 and 26% higher fold-activation for MR (Val-180)1compared to wild type MR. Corticosterone and 11-deoxycorticosterone had slightly lower EC50s and about 30% higher fold-activation for MR (Val-180) than for wild-type MR. Spironolactone and progesterone, which inhibit wild-type MR also inhibited aldosterone activation of MR (Val-180) [Figure 3].

The weaker EC50 and higher fold-activation of cortisol for MR (Val-180) by themselves may be insufficient to support the profound biological response to stress and depression in people with MR (Val-180) [20,45,46,69,70]. We intend to investigate other mechanism(s) that may contribute to novel physiological activity of MR (Val-180). One possibility arises from the different response to cortisol between MR (Val-180) and wild-type MR in the presence of either TAT3 or MMTV [40,74], and the similar finding by DeRijk et al. [43] in the response to cortisol of MR (Val-180) in cells with either TAT3 or MMTV promoters. It may be that other promoters in the brain exert larger differences between MR (Val-180) and wild-type MR in transcriptional activation by cortisol and other corticosteroids, which may be physiologically important.

Another possible mechanism is regulation of cortisol-mediated transcription of MR (Val-180) by phosphorylation of one or more serine residues near codon 180 the NTD on the MR [54–58]. Indeed, Shibata et al [54,57], Kino et al. [58] and Faresse et al. [75] have identified several phosphorylation sites in the NTD of human MR. These sites include Ser-129, Ser-183, Ser-196, Ser-227, Ser238, Ser-250 and Ser-263. Substitution of one or more serines in MR (Val-180) with alanine or aspartic acid would determine whether phosphorylation is important in corticosteroid activation of MR (Val-180).

We also will investigate the effect of the formation of heterodimers between MR (Val-180) and the GR [59–64,76] on activation of MR (Val-180), which is relevant to MR activity in the brain based on research by Mifsud and Ruel [61], who found that stress promotes MR-GR heterodimers in the hippocampus. Promotion by stress of heterodimers between MR (Val-180) and the GR may be important in the effects of MR (Val-180) on stress and depression.

**Materials & Methods**

**Construction of plasmid vectors**

Full-length mineralocorticoid receptor (MR) of human as registered in Genbank (accession number: XP_054206038) was used as wild-type human MR. MR rs5522 was constructed by the replacement of isoleucine-180 by valine using KOD-Plus-mutagenesis kit (TOYOBO). The nucleic acid sequences of all constructs were verified by sequencing.

**Chemical reagents**

Cortisol, corticosterone, 11-deoxycorticosterone, aldosterone, progesterone and spironolactone were purchased from Sigma-Aldrich. For reporter gene assays, all hormones were dissolved in dimethyl-sulfoxide (DMSO); the final DMSO concentration in the culture medium did not exceed 0.1%.

**Transactivation assays and statistical analyses**

Transfection and reporter assays were carried out in HEK293 cells, as described previously [29]. HEK293 cells were seeded in 24-well plates at $5 \times 10^4$ cells/well in phenol-red free Dulbecco's modified Eagle's medium supplemented with 10% charcoal-stripped fetal calf serum (HyClone). After 24 hours, the cells were transfected with 100 ng of receptor gene, reporter gene containing the *Photinus pyralis* luciferase gene and pRL-tk, as an internal control to normalize for variation in transfection efficiency; pRL-tk contains the *Renilla reniformis* luciferase gene with the herpes simplex virus thymidine kinase promoter. Each assay had a similar number of cells, and assays were done with the same batch of cells in each experiment. All experiments were performed in triplicate. Promoter activity was calculated as firefly (*P. pyralis*)-lucifease activity/sea pansy (*R. reniformis*)-lucifease activity. The values shown are mean ± SEM from three separate experiments, and dose-response data, which were used to calculate the half maximal response (EC50) for each steroid, were analyzed using GraphPad Prism.

**Antibodies and Western blotting of FLAG-tagged MR**

Primary antibodies for Western blotting were purchased and used at the indicated dilutions: anti-FLAG (clone #1E6, FUJIFILM Wako Pure Chemical Corporation, 1:750), and anti-betaActin (clone # 6D1, Medical & Biological Laboratories Co., LTD., 1:3000). Cells transfected as described in "Transactivation assays and statistical analysis" were directly lysed with sodium dodecyl sulfate (SDS) sample buffer and subjected to SDS-PAGE on a 10% polyacrylamide gel and transferred to an Immobilon membrane. The membrane was blocked with 10% skim milk overnight at 4 °C and then incubated with a primary antibody for 4 hours at room temperature. The membrane was washed with TTBS (0.1% Tween-20 in Tris-buffered saline) for at three times for 10 min and incubated with AP-conjugated secondary antibody (COSMO BIO Co. Ltd.) for 1 hour at room temperature. After the final wash with TTBS, the membrane was treated with NBT-BCIP solution). The gels were also stained with Coomassie Blue.

**Funding:** This work was supported by Grants-in-Aid for Scientific Research from the Ministry of Education, Culture, Sports, Science and Technology of Japan (23K05839) to Y.K., and the Takeda Science Foundation to Y.K.

**Competing interests:** The authors have declared that no competing interests exist.

**Contributions**.

Yoshinao Katsu: Investigation, Conceptualization, Supervision, Formal Analysis, Writing – original draft, Writing – review & editing.

Jiawen Zhang: Data curation, Investigation, Methodology.

Ya Ao: Data curation, Methodology.

Michael E. Baker: Conceptualization; Formal analysis, Supervision, Writing – original draft, Writing – review & editing.

# References

1. Arriza JL, Weinberger C, Cerelli G, et al. Cloning of human mineralocorticoid receptor complementary DNA: structural and functional kinship with the glucocorticoid receptor. Science. 1987;237(4812):268-275. doi:10.1126/science.3037703.

2. Evans RM. The steroid and thyroid hormone receptor superfamily. Science. 1988;240(4854):889-895. doi:10.1126/science.3283939.

3. Bridgham JT, Eick GN, Larroux C, et al. Protein evolution by molecular tinkering: diversification of the nuclear receptor superfamily from a ligand-dependent ancestor. PLoS Biol. 2010;8(10):e1000497. Published 2010 Oct 5. doi:10.1371/journal.pbio.1000497.

4. Baker ME, Nelson DR, Studer RA. Origin of the response to adrenal and sex steroids: Roles of promiscuity and co-evolution of enzymes and steroid receptors. J Steroid Biochem Mol Biol. 2015;151:12-24. doi:10.1016/j.jsbmb.2014.10.020.

5. Baker ME. Steroid receptors and vertebrate evolution. Mol Cell Endocrinol. 2019;496:110526. doi:10.1016/j.mce.2019.110526.

6. Evans RM, Mangelsdorf DJ. Nuclear Receptors, RXR, and the Big Bang. Cell. 2014 Mar 27;157(1):255-66. doi: 10.1016/j.cell.2014.03.012. PMID: 24679540; PMCID: PMC4029515.

7. Rossier BC, Baker ME, Studer RA. Epithelial sodium transport and its control by aldosterone: the story of our internal environment revisited. Physiol Rev. 2015;95(1):297-340. doi:10.1152/physrev.00011.2014.

8. Hanukoglu I, Hanukoglu A. Epithelial sodium channel (ENaC) family: Phylogeny, structure-function, tissue distribution, and associated inherited diseases. Gene. 2016;579(2):95-132. doi:10.1016/j.gene.2015.12.061.

9. Lifton RP, Gharavi AG, Geller DS. Molecular mechanisms of human hypertension. Cell. 2001;104(4):545-556. doi:10.1016/s0092-8674(01)00241-0.

10. Shibata S. 30 YEARS OF THE MINERALOCORTICOID RECEPTOR: Mineralocorticoid receptor and NaCl transport mechanisms in the renal distal nephron. J Endocrinol. 2017;234(1):T35-T47. doi:10.1530/JOE-16-0669.

11. Grossmann C, Almeida-Prieto B, Nolze A, Alvarez de la Rosa D. Structural and molecular determinants of mineralocorticoid receptor signalling. Br J Pharmacol. 2021 Nov 22. doi: 10.1111/bph.15746. Epub ahead of print. PMID: 34811739.

12. Fernandes-Rosa FL, Boulkroun S, Fedlaoui B, Hureaux M, Travers-Allard S, Drossart T, Favier J, Zennaro MC. New advances in endocrine hypertension: from genes to biomarkers.15

Kidney Int. 2023 Mar;103(3):485-500. doi: 10.1016/j.kint.2022.12.021. Epub 2023 Jan 13. PMID: 36646167.

13. Hawkins UA, Gomez-Sanchez EP, Gomez-Sanchez CM, Gomez-Sanchez CE. The ubiquitous mineralocorticoid receptor: clinical implications. Curr Hypertens Rep. 2012;14(6):573-580. doi:10.1007/s11906-012-0297-0.

14. Jaisser F, Farman N. Emerging Roles of the Mineralocorticoid Receptor in Pathology: Toward New Paradigms in Clinical Pharmacology. Pharmacol Rev. 2016;68(1):49-75. doi:10.1124/pr.115.011106.

15. Funder J. 30 YEARS OF THE MINERALOCORTICOID RECEPTOR: Mineralocorticoid receptor activation and specificity-conferring mechanisms: a brief history. J Endocrinol. 2017 Jul;234(1):T17-T21. doi: 10.1530/JOE-17-0119. Epub 2017 May 22. PMID: 28533421.

16. Gomez-Sanchez EP. Brain mineralocorticoid receptors in cognition and cardiovascular homeostasis. Steroids. 2014 Dec;91:20-31. doi: 10.1016/j.steroids.2014.08.014. PMID: 25173821; PMCID: PMC4302001.

17. De Kloet ER, Vreugdenhil E, Oitzl MS, Joëls M. Brain corticosteroid receptor balance in health and disease. Endocr Rev. 1998 Jun;19(3):269-301. doi: 10.1210/edrv.19.3.0331. PMID: 9626555.

18. Arriza JL, Simerly RB, Swanson LW, Evans RM. The neuronal mineralocorticoid receptor as a mediator of glucocorticoid response. Neuron. 1988 Nov;1(9):887-900. doi: 10.1016/0896-6273(88)90136-5. PMID: 2856104.

19. de Kloet ER, Joëls M. Brain mineralocorticoid receptor function in control of salt balance and stress-adaptation. Physiol Behav. 2017 Sep 1;178:13-20. doi: 10.1016/j.physbeh.2016.12.045. Epub 2017 Jan 13. PMID: 28089704.

20. Paul SN, Wingenfeld K, Otte C, Meijer OC. Brain mineralocorticoid receptor in health and disease: From molecular signalling to cognitive and emotional function. Br J Pharmacol. 2022 Jul;179(13):3205-3219. doi: 10.1111/bph.15835. Epub 2022 Apr 7. PMID: 35297038; PMCID: PMC9323486.

21. Ibarrola J, Jaffe IZ. The Mineralocorticoid Receptor in the Vasculature: Friend or Foe? Annu Rev Physiol. 2024 Feb 12;86:49-70. doi: 10.1146/annurev-physiol-042022-015223. Epub 2023 Oct 3. PMID: 37788489.

22. de Kloet ER. Brain mineralocorticoid and glucocorticoid receptor balance in neuroendocrine regulation and stress-related psychiatric etiopathologies. Curr Opin Endocr Metab Res. 2022 Jun;24:100352. doi: 10.1016/j.coemr.2022.100352. PMID: 38037568; PMCID: PMC10687720.

23. Gaudenzi C, Mifsud KR, Reul JMHM. Insights into isoform-specific mineralocorticoid receptor action in the hippocampus. J Endocrinol. 2023 Jul 12;258(2):e220293. doi: 10.1530/JOE-22-0293. PMID: 37235709.

24. Yang J, Fuller PJ. Interactions of the mineralocorticoid receptor--within and without. Mol Cell Endocrinol. 2012 Mar 24;350(2):196-205. doi: 10.1016/j.mce.2011.07.001. Epub 2011 Jul 18. PMID: 21784126.

25. Beato M, Arnemann J, Chalepakis G, Slater E, Willmann T. Gene regulation by steroid hormones. J Steroid Biochem. 1987;27(1-3):9-14. doi: 10.1016/0022-4731(87)90288-3. PMID: 2826895.

26. Baker ME, Katsu Y. 30 YEARS OF THE MINERALOCORTICOID RECEPTOR: Evolution of the mineralocorticoid receptor: sequence, structure and function. J Endocrinol. 2017;234(1):T1-T16. doi:10.1530/JOE-16-0661.

27. Rogerson FM, Fuller PJ. Interdomain interactions in the mineralocorticoid receptor. Mol Cell Endocrinol. 2003 Feb 28;200(1-2):45-55. doi: 10.1016/s0303-7207(02)00413-6. PMID: 12644298.

28. Rupprecht R, Arriza JL, Spengler D, et al. Transactivation and synergistic properties of the mineralocorticoid receptor: relationship to the glucocorticoid receptor. Mol Endocrinol. 1993;7(4):597-603. doi:10.1210/mend.7.4.8388999.

29. Katsu Y, Oka K, Baker ME. Evolution of human, chicken, alligator, frog, and zebrafish mineralocorticoid receptors: Allosteric influence on steroid specificity. Sci Signal. 2018 Jul 3;11(537):eaao1520. doi: 10.1126/scisignal.aao1520. PMID: 29970600.

30. Funder JW. Glucocorticoid and mineralocorticoid receptors: biology and clinical relevance. Annu Rev Med. 1997;48:231-40. doi: 10.1146/annurev.med.48.1.231. PMID: 9046958.

31. Hellal-Levy C, Couette B, Fagart J, Souque A, Gomez-Sanchez C, Rafestin-Oblin M. Specific hydroxylations determine selective corticosteroid recognition by human glucocorticoid and mineralocorticoid receptors. FEBS Lett. 1999;464(1-2):9-13. doi:10.1016/s0014-5793(99)01667-1.

32. Grossmann C, Scholz T, Rochel M, Bumke-Vogt C, Oelkers W, Pfeiffer AF, Diederich S, Bahr V. Transactivation via the human glucocorticoid and mineralocorticoid receptor by therapeutically used steroids in CV-1 cells: a comparison of their glucocorticoid and mineralocorticoid properties. Eur J Endocrinol. 2004 Sep;151(3):397-406. doi: 10.1530/eje.0.1510397. PMID: 15362971.

33. Shimeld SM, Donoghue PC. Evolutionary crossroads in developmental biology: cyclostomes (lamprey and hagfish). Development. 2012 Jun;139(12):2091-9. doi: 10.1242/dev.074716. PMID: 22619386.

34. Close DA, Yun SS, McCormick SD, Wildbill AJ, Li W. 11-deoxycortisol is a corticosteroid hormone in the lamprey. Proc Natl Acad Sci U S A. 2010 Aug 3;107(31):13942-7. doi: 10.1073/pnas.0914026107. Epub 2010 Jul 19. PMID: 20643930; PMCID: PMC2922276.

35. Thornton JW. Evolution of vertebrate steroid receptors from an ancestral estrogen receptor by ligand exploitation and serial genome expansions. Proc Natl Acad Sci U S A. 2001 May 8;98(10):5671-6. doi: 10.1073/pnas.091553298. Epub 2001 May 1. PMID: 11331759; PMCID: PMC33271.

36. Smith JJ, Kuraku S, Holt C, Sauka-Spengler T, Jiang N, Campbell MS, Yandell MD, Manousaki T, Meyer A, Bloom OE, Morgan JR, Buxbaum JD, Sachidanandam R, Sims C, Garruss AS, Cook M, Krumlauf R, Wiedemann LM, Sower SA, Decatur WA, Hall JA, Amemiya CT, Saha NR, Buckley KM, Rast JP, Das S, Hirano M, McCurley N, Guo P, Rohner N, Tabin CJ, Piccinelli P, Elgar G, Ruffier M, Aken BL, Searle SM, Muffato M, Pignatelli M, Herrero J, Jones M, Brown CT, Chung-Davidson YW, Nanlohy KG, Libants SV, Yeh CY, McCauley DW, Langeland JA, Pancer Z, Fritzsch B, de Jong PJ, Zhu B, Fulton LL, Theising B, Flicek P, Bronner ME, Warren WC, Clifton SW, Wilson RK, Li W. Sequencing of the sea lamprey (Petromyzon marinus) genome provides insights into vertebrate evolution. Nat Genet. 2013 Apr;45(4):415-21, 421e1-2. doi: 10.1038/ng.2568. Epub 2013 Feb 24. PMID: 23435085; PMCID: PMC3709584.

37. York JR, McCauley DW. The origin and evolution of vertebrate neural crest cells. Open Biol. 2020 Jan;10(1):190285. doi: 10.1098/rsob.190285. Epub 2020 Jan 29. PMID: 31992146; PMCID: PMC7014683.

38. Kuratani S. Evo-devo studies of cyclostomes and the origin and evolution of jawed vertebrates. Curr Top Dev Biol. 2021;141:207-239. doi: 10.1016/bs.ctdb.2020.11.011. Epub 2020 Dec 13. PMID: 33602489.

39. Janvier P. microRNAs revive old views about jawless vertebrate divergence and evolution. Proc Natl Acad Sci U S A. 2010 Nov 9;107(45):19137-8. doi: 10.1073/pnas.1014583107. Epub 2010 Nov 1. PMID: 21041649; PMCID: PMC2984170.

40. Katsu Y, Zhang J, Baker ME. Lysine-Cysteine-Serine-Tryptophan inserted into the DNA-binding domain of human mineralocorticoid receptor increases transcriptional activation by aldosterone. J Steroid Biochem Mol Biol. 2024 Oct;243:106548. doi: 10.1016/j.jsbmb.2024.106548. Epub 2024 May 29. PMID: 38821293.

41. Arai K, Tsigos C, Suzuki Y, Irony I, Karl M, Listwak S, Chrousos GP. Physiological and molecular aspects of mineralocorticoid receptor action in pseudohypoaldosteronism: a responsiveness test and therapy. J Clin Endocrinol Metab. 1994 Oct;79(4):1019-23. doi: 10.1210/jcem.79.4.7962269. PMID: 7962269.

42. Arai K, Nakagomi Y, Iketani M, Shimura Y, Amemiya S, Ohyama K, Shibasaki T. Functional polymorphisms in the mineralocorticoid receptor and amiloride-sensitive sodium channel genes in a patient with sporadic pseudohypoaldosteronism. Hum Genet. 2003 Jan;112(1):91-7. doi: 10.1007/s00439-002-0855-7. Epub 2002 Oct 25. PMID: 12483305.

43. DeRijk RH, Wüst S, Meijer OC, Zennaro MC, Federenko IS, Hellhammer DH, Giacchetti G, Vreugdenhil E, Zitman FG, de Kloet ER. A common polymorphism in the mineralocorticoid receptor modulates stress responsiveness. J Clin Endocrinol Metab. 2006 Dec;91(12):5083-9. doi: 10.1210/jc.2006-0915. Epub 2006 Oct 3. PMID: 17018659.

44. Derijk RH, de Kloet ER. Corticosteroid receptor polymorphisms: determinants of vulnerability and resilience. Eur J Pharmacol. 2008 Apr 7;583(2-3):303-11. doi: 10.1016/j.ejphar.2007.11.072. Epub 2008 Jan 30. PMID: 18321483.

45. de Kloet ER, Otte C, Kumsta R, Kok L, Hillegers MH, Hasselmann H, Kliegel D, Joëls M. Stress and Depression: a Crucial Role of the Mineralocorticoid Receptor. J Neuroendocrinol. 2016 Aug;28(8). doi: 10.1111/jne.12379. PMID: 26970338.

46. Bogdan R, Perlis RH, Fagerness J, Pizzagalli DA. The impact of mineralocorticoid receptor ISO/VAL genotype (rs5522) and stress on reward learning. Genes Brain Behav. 2010 Aug;9(6):658-67. doi: 10.1111/j.1601-183X.2010.00600.x. Epub 2010 Jun 7. PMID: 20528958; PMCID: PMC2921022.

47. Kuningas M, de Rijk RH, Westendorp RG, Jolles J, Slagboom PE, van Heemst D. Mental performance in old age dependent on cortisol and genetic variance in the mineralocorticoid and glucocorticoid receptors. Neuropsychopharmacology. 2007 Jun;32(6):1295-301. doi: 10.1038/sj.npp.1301260. Epub 2006 Nov 29. PMID: 17133261.

48. van Leeuwen N, Bellingrath S, de Kloet ER, Zitman FG, DeRijk RH, Kudielka BM, Wüst S. Human mineralocorticoid receptor (MR) gene haplotypes modulate MR expression and transactivation: implication for the stress response. Psychoneuroendocrinology. 2011 Jun;36(5):699-709. doi: 10.1016/j.psyneuen.2010.10.003. Epub 2010 Nov 20. PMID: 21095064.

49. Geller DS, Farhi A, Pinkerton N, et al. Activating mineralocorticoid receptor mutation in hypertension exacerbated by pregnancy. Science. 2000;289(5476):119-123. doi:10.1126/science.289.5476.119.

50. Kolkhof P, Bärfacker L. 30 YEARS OF THE MINERALOCORTICOID RECEPTOR: Mineralocorticoid receptor antagonists: 60 years of research and development. J Endocrinol. 2017 Jul;234(1):T125-T140. doi: 10.1530/JOE-16-0600. PMID: 28634268; PMCID: PMC5488394.

51. Iñiguez-Lluhí JA, Pearce D. A common motif within the negative regulatory regions of multiple factors inhibits their transcriptional synergy. Mol Cell Biol. 2000 Aug;20(16):6040-50. doi: 10.1128/mcb.20.16.6040-6050.2000. PMID: 10913186; PMCID: PMC86080.

52. Lombès M, Binart N, Oblin ME, Joulin V, Baulieu EE. Characterization of the interaction of the human mineralocorticosteroid receptor with hormone response elements. Biochem J. 1993 Jun 1;292 ( Pt 2)(Pt 2):577-83. doi: 10.1042/bj2920577. PMID: 8389140; PMCID: PMC1134249.

53. Cato AC, Skroch P, Weinmann J, Butkeraitis P, Ponta H. DNA sequences outside the receptor-binding sites differently modulate the responsiveness of the mouse mammary tumour virus promoter to various steroid hormones. EMBO J. 1988 May;7(5):1403-10. PMID: 2842149; PMCID: PMC458390.

54. Shibata S, Ishizawa K, Wang Q, Xu N, Fujita T, Uchida S, Lifton RP. ULK1 Phosphorylates and Regulates Mineralocorticoid Receptor. Cell Rep. 2018 Jul 17;24(3):569-576. doi: 10.1016/j.celrep.2018.06.072. PMID: 30021155.

55. Faresse N. Post-translational modifications of the mineralocorticoid receptor: How to dress the receptor according to the circumstances? J Steroid Biochem Mol Biol. 2014 Sep;143:334-42. doi: 10.1016/j.jsbmb.2014.04.015. Epub 2014 May 10. PMID: 24820770.

56. Gadasheva Y, Nolze A, Grossmann C. Posttranslational Modifications of the Mineralocorticoid Receptor and Cardiovascular Aging. Front Mol Biosci. 2021 May 28;8:667990. doi: 10.3389/fmolb.2021.667990. PMID: 34124152; PMCID: PMC8193679.

57. Shibata S, Rinehart J, Zhang J, Moeckel G, Castañeda-Bueno M, Stiegler AL, Boggon TJ, Gamba G, Lifton RP. Mineralocorticoid receptor phosphorylation regulates ligand binding and renal response to volume depletion and hyperkalemia. Cell Metab. 2013 Nov 5;18(5):660-71. doi: 10.1016/j.cmet.2013.10.005. PMID: 24206662; PMCID: PMC3909709.

58. Kino T, Jaffe H, Amin ND, Chakrabarti M, Zheng YL, Chrousos GP, Pant HC. Cyclin-dependent kinase 5 modulates the transcriptional activity of the mineralocorticoid receptor and regulates expression of brain-derived neurotrophic factor. Mol Endocrinol. 2010 May;24(5):941-52. doi: 10.1210/me.2009-0395. Epub 2010 Mar 31. PMID: 20357208; PMCID: PMC2870940.

59. Liu W, Wang J, Sauter NK, Pearce D. Steroid receptor heterodimerization demonstrated in vitro and in vivo. Proc Natl Acad Sci U S A. 1995 Dec 19;92(26):12480-4. doi: 10.1073/pnas.92.26.12480. PMID: 8618925; PMCID: PMC40381.

60. Trapp T, Holsboer F. Heterodimerization between mineralocorticoid and glucocorticoid receptors increases the functional diversity of corticosteroid action. Trends Pharmacol Sci. 1996 Apr;17(4):145-9. doi: 10.1016/0165-6147(96)81590-2. PMID: 8984741.

61. Mifsud KR, Reul JM. Acute stress enhances heterodimerization and binding of corticosteroid receptors at glucocorticoid target genes in the hippocampus. Proc Natl Acad Sci U S A. 2016 Oct 4;113(40):11336-11341. doi: 10.1073/pnas.1605246113. Epub 2016 Sep 21. PMID: 27655894; PMCID: PMC5056104.

62. Pooley JR, Rivers CA, Kilcooley MT, Paul SN, Cavga AD, Kershaw YM, Muratcioglu S, Gursoy A, Keskin O, Lightman SL. Beyond the heterodimer model for mineralocorticoid and glucocorticoid receptor interactions in nuclei and at DNA. PLoS One. 2020 Jan 10;15(1):e0227520. doi: 10.1371/journal.pone.0227520. PMID: 31923266; PMCID: PMC6953809.

63. Kiilerich P, Triqueneaux G, Christensen NM, et al. Interaction between the trout mineralocorticoid and glucocorticoid receptors in vitro. J Mol Endocrinol. 2015;55(1):55-68. doi:10.1530/JME-15-0002.

64. Alvarez de la Rosa D, Ramos-Hernández Z, Weller-Pérez J, Johnson TA, Hager GL. The impact of mineralocorticoid and glucocorticoid receptor interaction on corticosteroid transcriptional outcomes. Mol Cell Endocrinol. 2024 Dec 1;594:112389. doi: 10.1016/j.mce.2024.112389. Epub 2024 Oct 17. PMID: 39423940.

65. Meijsing SH, Pufall MA, So AY, Bates DL, Chen L, Yamamoto KR. DNA binding site sequence directs glucocorticoid receptor structure and activity. Science. 2009 Apr 17;324(5925):407-10. doi: 10.1126/science.1164265. PMID: 19372434; PMCID: PMC2777810.

66. Chapman K, Holmes M, Seckl J. 11β-hydroxysteroid dehydrogenases: intracellular gate-keepers of tissue glucocorticoid action. Physiol Rev. 2013 Jul;93(3):1139-206. doi: 10.1152/physrev.00020.2012. PMID: 23899562; PMCID: PMC3962546.

67. Frey FJ, Odermatt A, Frey BM. Glucocorticoid-Mediated Mineralocorticoid Receptor Activation and Hypertension. Curr Opin Nephrol Hypertens. 2004 Jul;13(4):451-8. Doi: 10.1097/01.Mnh.0000133976.32559.B0. PMID: 15199296.

68. Funder JW. Apparent mineralocorticoid excess. J Steroid Biochem Mol Biol. 2017 Jan;165(Pt A):151-153. doi: 10.1016/j.jsbmb.2016.03.010. Epub 2016 Mar 5. PMID: 26956190.

69. Klok MD, Alt SR, Irurzun Lafitte AJ, Turner JD, Lakke EA, Huitinga I, Muller CP, Zitman FG, de Kloet ER, Derijk RH. Decreased expression of mineralocorticoid receptor mRNA and its splice variants in postmortem brain regions of patients with major depressive disorder. J Psychiatr Res. 2011 Jul;45(7):871-8. doi: 10.1016/j.jpsychires.2010.12.002. Epub 2010 Dec 30. PMID: 21195417.

70. Medina A, Seasholtz AF, Sharma V, Burke S, Bunney W Jr, Myers RM, Schatzberg A, Akil H, Watson SJ. Glucocorticoid and mineralocorticoid receptor expression in the human hippocampus in major depressive disorder. J Psychiatr Res. 2013 Mar;47(3):307-14. doi: 10.1016/j.jpsychires.2012.11.002. Epub 2012 Dec 6. PMID: 23219281; PMCID: PMC4248661.

71. Klok MD, Giltay EJ, Van der Does AJ, Geleijnse JM, Antypa N, Penninx BW, de Geus EJ, Willemsen G, Boomsma DI, van Leeuwen N, Zitman FG, de Kloet ER, DeRijk RH. A common and functional mineralocorticoid receptor haplotype enhances optimism and protects against depression in females. Transl Psychiatry. 2011 Dec 13;1(12):e62. doi: 10.1038/tp.2011.59. PMID: 22832354; PMCID: PMC3309494.

72. Fuse H, Kitagawa H, Kato S. Characterization of transactivational property and coactivator mediation of rat mineralocorticoid receptor activation function-1 (AF-1). Mol Endocrinol. 2000 Jun;14(6):889-99. doi: 10.1210/mend.14.6.0467. PMID: 10847590.

73. Fischer K, Kelly SM, Watt K, Price NC, McEwan IJ. Conformation of the mineralocorticoid receptor N-terminal domain: evidence for induced and stable structure. Mol Endocrinol. 2010 Oct;24(10):1935-48. doi: 10.1210/me.2010-0005. Epub 2010 Aug 4. PMID: 20685853; PMCID: PMC5417395.

74. Katsu Y, Oana S, Lin X, Hyodo S, Baker ME. Aldosterone and dexamethasone activate African lungfish mineralocorticoid receptor: Increased activation after removal of the amino-terminal domain. J Steroid Biochem Mol Biol. 2022 Jan;215:106024. doi: 10.1016/j.jsbmb.2021.106024. Epub 2021 Nov 10. PMID: 34774724.

75. Faresse N, Vitagliano JJ, Staub O. Differential ubiquitylation of the mineralocorticoid receptor is regulated by phosphorylation. FASEB J. 2012 Oct;26(10):4373-82. doi: 10.1096/fj.12-209924. Epub 2012 Jul 13. PMID: 22798426.

76. Johnson TA, Fettweis G, Wagh K, Ceacero-Heras D, Krishnamurthy M, Sánchez de Medina F, Martínez-Augustin O, Upadhyaya A, Hager GL, Alvarez de la Rosa D. The glucocorticoid receptor potentiates aldosterone-induced transcription by the mineralocorticoid receptor. Proc Natl Acad Sci U S A. 2024 Nov 19;121(47):e2413737121. doi: 10.1073/pnas.2413737121. Epub 2024 Nov 14. PMID: 39541347; PMCID: PMC11588051.